\documentclass{article}
\usepackage{spconf,amsmath,graphicx,hyperref}

\usepackage{pifont}
\newcommand{\cmark}{\ding{51}}
\newcommand{\xmark}{\ding{55}}

\usepackage{booktabs}
\usepackage{tablefootnote}
\usepackage{array}
\usepackage{multirow}
\usepackage{enumitem}

\usepackage{cite}

\title{PAS-SE: Personalized Auxiliary-Sensor Speech Enhancement\\for Voice Pickup in Hearables}

\name{Mattes Ohlenbusch$^{1,2}$\sthanks{This work was done during an internship at Bose.} \qquad Mikolaj Kegler$^1$ \qquad Marko Stamenovic$^1$}
\address{
\small
$^1$Bose Corporation, USA\\
\small
$^2$Fraunhofer Institute for Digital Media Technology IDMT, Oldenburg Branch for Hearing, Speech and Audio Technology HSA, Germany
}
\begin{document}
\ninept
\maketitle

\begin{abstract} 
Speech enhancement for voice pickup in hearables aims to improve the user’s voice by suppressing noise and interfering talkers, while maintaining own-voice quality. 
For single-channel methods, it is particularly challenging to distinguish the target from interfering talkers without additional context. 
In this paper, we compare two strategies to resolve this ambiguity:
personalized speech enhancement (PSE), which uses enrollment utterances to represent the target, and auxiliary‑sensor speech enhancement (AS-SE), which uses in‑ear microphones as additional input.
We evaluate the strategies on two public datasets, employing different auxiliary sensor arrays, to investigate their cross-dataset generalization.
We propose training‑time augmentations to facilitate cross-dataset generalization of AS-SE systems. We also show that combining PSE and AS-SE (PAS-SE) provides complementary performance benefits, especially when enrollment speech is recorded with the in‑ear microphone.
We further demonstrate that PAS-SE personalized with noisy in-ear enrollments maintains performance benefits over the AS-SE system.
\end{abstract}


\section{Introduction}
Hearable devices with one or more microphones can be used to capture the user’s own voice (i.e., target speech), but due to environmental noise and interfering talkers, the captured speech signal needs to be enhanced to facilitate communication. Although single-channel speech enhancement (SE) approaches based on deep neural networks can reduce environmental noise~\cite{braun_towards_2021, rong_gtcrn_2024, fedorov2020tinylstms,nathoo2024two}, they tend to struggle with removing interfering talkers from the mixture while preserving the user's voice.
Modern hearables often include microphones placed inside or near the user's occluded ear canal, typically employed in active noise reduction systems~\cite{ChangListening2016}. These in-ear microphones are acoustically shielded from environmental noise and interfering talkers by the device. At the same time, the user's voice is also picked up by the in-ear microphone, predominantly through body conduction~\cite{bouserhal_-ear_2019, ohlenbusch_modeling_2023}. These two effects lead to a substantial benefit in terms of signal-to-noise ratio (SNR) of the user's voice at the in-ear microphone. However, this signal cannot be directly used for voice communication due to time- and user-varying band-limitation, nonlinearities and distortions
introduced by body conduction~\cite{bouserhal_-ear_2019, ohlenbusch_modeling_2023}, and undesired additive body-produced noises~\cite{gaardbaek_origin_2025}. 
Other auxiliary body-conduction sensors, such as accelerometers, have similar properties and trade-offs as in-ear microphones~\cite{tagliasacchi20_interspeech}.

Due to their benefits, auxiliary sensors have been employed for own-voice speech enhancement~\cite{tagliasacchi20_interspeech, yu_time-domain_2020, wang_fusing_2022, wang_multi-modal_2022, edraki_speaker_2024}. 
The use of an auxiliary sensor as an additional input can substantially improve performance, especially in challenging SNRs~\cite{ohlenbusch_multi-microphone_2024} and scenarios with interfering talkers~\cite{tagliasacchi20_interspeech}.
Notably, in~\cite{tagliasacchi20_interspeech} it was found that using a body-conduction sensor as additional input facilitated interferer suppression when tested with simulated data, assuming no noise and interferer transmission to the auxiliary sensor. We broadly refer to this approach, where the system uses both an outer microphone and an auxiliary sensor as input, as auxiliary sensor speech enhancement (AS-SE).
Another way to extract speech from a target talker in a single-channel system is to condition the system with a feature vector obtained from enrollment utterances of the target talker~\cite{delcroix_tdspeakerbeam_2020, Xu_Spex_2020, eskimez_personalized_2022, ZmolikovaOverview2023, sinha2024variants}, commonly referred to as target speaker extraction (TSE). 
In this work, we consider a special case of TSE, personalized speech enhancement (PSE), where the target talker is always the device user.
Using the enrollment utterance of the user, the PSE system can resolve the ambiguity between target and interfering talker.

It should be noted that the methods mentioned above each come with their own unique trade-offs.
PSE typically requires enrollment speech, which implies an additional setup procedure for the user prior to using the system. 
In contrast, AS-SE systems can be readily used by any user without an additional setup procedure.
However, the resulting system may not generalize across different devices due to unique array properties or acoustic design of the hearable device. 

A systematic evaluation or integration of PSE and AS-SE methods has yet to be carried out. 
In this paper, we:
\begin{enumerate}[noitemsep]
    \item Benchmark PSE and AS-SE performance by systematically evaluating their denoising capabilities, suppression of interfering talkers, and generalization across datasets.
    \item Explore and ablate different training-time augmentation configurations for training AS-SE systems with limited pre-recorded in-ear signals to facilitate interfering talker suppression and cross-dataset generalization.
    \item Personalize AS-SE through enrollment-based conditioning, investigate the benefits of auxiliary-sensor enrollment to develop personalized AS-SE (PAS-SE) systems, and analyze their robustness to noise in enrollment utterances.
\end{enumerate}

Experimental evaluations carried out on two openly available in-ear microphone datasets demonstrate that interfering speech can be considerably reduced with either PSE or AS-SE, while a combination of both, which we term personalized auxiliary-sensor speech enhancement (PAS-SE), leads to further improvements. 
In particular, PAS-SE personalized with in-ear enrollments leads to the best results, generalizing both within and across datasets tested. Code and audio examples are available online\footnote{\href{https://bose.github.io/passe/}{https://bose.github.io/passe/}.
}.

\section{Signal model}
We consider a hearable device equipped with an outer microphone and an in-ear microphone. 
The signals are denoted by subscripts $o$ for the outer microphone~(OM) and $i$ for the in-ear microphone~(IM). 
In the short-time Fourier transform (STFT) domain, $S_o(k,l)$ and $S_i(k,l)$ denote the own voice signals of the user (i.e., target speech) at each microphone, where $k$ and $l$ denote the frequency  and the frame index.
The outer and in-ear microphone signals are given by 
\begin{equation}
    Y_{\{o,i\}}(k,l) =  S_{\{o,i\}}(k,l) + N_{\{o,i\}}(k,l) + V_{\{o,i\}}(k,l), \label{eq:sigmodel_outer_inear}
\end{equation}
respectively, where the noise components are denoted by $N_o(k,l)$ and $N_i(k,l)$, and the interfering talker components are denoted by $V_o(k,l)$ and $V_i(k,l)$. 

\vspace{-0.3cm}
\section{System architecture}
\begin{figure}
    \centering
    \includegraphics
    [width=0.9\columnwidth]
    {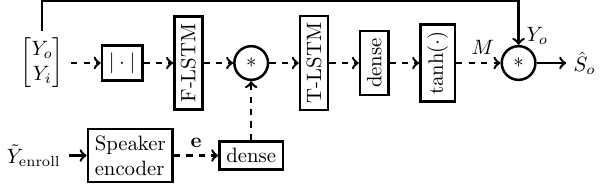}
    \vspace{-0.3cm}
    \caption{PAS-SE system architecture based on FT-JNF~\cite{tesch_insights_2023}.
    The system is personalized using multiplicative conditioning with a feature vector $\mathbf{e}$ obtained from an enrollment utterance $\tilde{Y}_\text{enroll}$.}
    \label{fig:ftjnf_diagram}
\end{figure}

In this paper, we explore various SE, PSE, AS-SE, and PAS-SE systems by modifying the FT-JNF architecture proposed in~\cite{tesch_insights_2023}.
This architecture has previously been applied to the AS-SE task in~\cite{ohlenbusch_multi-microphone_2024, ohlenbusch_lowcomplexity_2025}.
Fig.~\ref{fig:ftjnf_diagram} shows the personalized FT-JNF variant used in this paper, 
which takes the magnitude of the noisy STFT coefficients of one or more microphones as input. 
The architecture consists of a unidirectional long short-term memory (LSTM) layer operating across the frequency dimension of the input (F-LSTM) with 512 hidden units, a causal unidirectional LSTM layer operating across the time dimension (T-LSTM) with 128 hidden units, a linear layer, and a $\tanh$ activation function.
The output is a magnitude mask $M(k,l)$, which is applied to the noisy outer microphone signal $Y_o(k,l)$, i.e. $\hat{S}_o(k,l) = M(k,l)\cdot Y_o(k,l)$. The SE system (with one input) has 1.384\,M parameters, and the AS-SE system has 1.386\,M parameters. 

We add a conditioning mechanism to the architecture to investigate personalization for our PSE and PAS-SE systems by introducing a speaker encoder branch based on the time-domain SpeakerBeam architecture~\cite{delcroix_tdspeakerbeam_2020} using the authors' implementation in the Asteroid toolkit~\cite{pariente20_interspeech}. 
An utterance $\tilde{Y}_\text{enroll}(k,l)$ from the same user, excluding the input utterance (either recorded at the outer or in-ear microphone), is used as input to the speaker encoder branch. 
The branch consists of a learnable filterbank encoder
, which is followed by a 1-D convolutional block
and temporal averaging to obtain a 128-dimensional speaker embedding vector $\mathbf{e}$.
A dense layer is then used to match the dimension of the output of the F-LSTM layer of the main enhancement network. The output of the F-LSTM layer is multiplied with the output of the dense layer from the speaker encoder branch. The speaker encoder consists of 1.810\,M parameters.

\section{Evaluation details}
\vspace{-0.3cm}
\subsection{Datasets}

The Vibravox Dataset~\cite{jhauret-et-al-2024-vibravox} was used for training and evaluation.
It contains approximately 33\,h of clean speech signals recorded from 198 talkers, for which we used the official split into training, validation, and test data of the \textit{speech\_clean} subset.
Recorded signals from the rigid earbud with an in-ear microphone~\cite{denk_one-size-fits-all_2019} (\textit{rigid in-ear}) were used for training. 
The close-talk microphone (a headset boom microphone) was selected in place of an outer microphone.
For environmental noise, we use the \textit{speechless\_noisy} subset of Vibravox. 

The Oldenburg dataset~\cite{ohlenbusch_modeling_2023} was used for the cross-dataset evaluation of the proposed systems (and training dataset-specific baselines).
The dataset consists of 306 utterances by 12, 2, and 4 different talkers across training, validation, and test, respectively.
The dataset was recorded using the same type of rigid earbud with an in-ear microphone~\cite{denk_one-size-fits-all_2019} as Vibravox.
Unlike Vibravox, which uses the close-talk mic, a microphone at the outer face of the device (most common in commodity hearables) is used as outer microphone. This induces a large difference in the auxiliary sensor array between the in-ear and outer microphones compared to the Vibravox setup.
For environmental noise, noise signals from the fifth DNS challenge~\cite{dubey_icassp_2024} were convolved with the individual impulse responses of the device user to obtain spatialized multi-channel noise signals. The impulse responses consisted of measurements from 8 evenly spaced loudspeakers in a circle around the user~\cite{ohlenbusch_multi-microphone_2024}. 
For interfering talkers, the same procedure is followed replacing noise signals with talkers from the Oldenburg training set other than the target talker.

The same procedure for mixing own voice, noise, and interferers is carried out for both datasets.
For each utterance during training, there is a 75\% probability of environmental noise being added.
If added, own voice and environmental noise are mixed to an SNR randomly selected from a uniform distribution between [-10, 10]\,dB (defined at the outer microphone). 
The corresponding scaling factor is applied to the in-ear microphone signals as well, to realistically preserve SNR differences between the noisy outer and acoustically shielded in-ear microphone.
The same procedure is applied to interferers, with an independent 75\% probability. This ensures the system is trained on a combination of user's clean speech, noisy speech, noisy speech with interferers, and speech with only interferers.
Since Vibravox does not contain an isolated interfering speaker partition, we randomly select from the set of speakers in the training partition disjoint from the target speaker as the interferer at each step.

\begin{table}
     \vspace{-6pt}
    \caption{Configurations of additive noise and interferer approximations at the outer and in-ear microphones for training AS-SE and PAS-SE systems. All training configurations include own voice recorded at both in-ear and outer microphones $S_{\{o,i\}}$.}
    \label{tab:approximations_noise_interferer}
    \centering
    \resizebox{0.9\columnwidth}{!}{
    \setlength{\tabcolsep}{0.5em}
    \begin{tabular}{l>{\centering\arraybackslash} p{1.5cm} >{\centering\arraybackslash}p{1.7cm} >{\centering\arraybackslash}p{1.5cm} >{\centering\arraybackslash}p{1.6cm} }
    \toprule
    & \multicolumn{2}{c}{\textbf{Outer microphone}} & \multicolumn{2}{c}{\textbf{In-ear microphone}} \\  
    &  Noise $N_o$ & Interferer $V_o$ & Noise $N_i$ & Interferer $V_i$ \\
    \midrule
        A \cite{tagliasacchi20_interspeech} & \cmark  & \cmark & \xmark & \xmark \\
        B \cite{ohlenbusch_multi-microphone_2024} & \cmark & \xmark & \cmark & \xmark \\
        C & \cmark & \cmark & \cmark & \xmark \\
        D & \cmark & \cmark & \cmark & $a\cdot V_o$ \\
    \bottomrule
    \end{tabular}
    }
\end{table}
\subsection{In-ear noise and interferer training configurations}
To facilitate training of AS-SE and PAS-SE systems without recorded interferer signals (as in Vibravox), we explore various configurations of incorporating in-ear signal noise and interferer components during training as shown in Table~\ref{tab:approximations_noise_interferer}. 
The components in~\eqref{eq:sigmodel_outer_inear} are either included using recorded signals (\cmark) or not (\xmark) during training, or the in-ear microphone component is approximated by the corresponding outer microphone component attenuated using a random factor $a \in 
[0.001, 1]$ from a uniform distribution.

In configuration (A), we consider training an AS-SE system by adding noise $N_o$ and interfering speech $V_o$ only to the outer microphone, but not to the auxiliary sensor signal as in~\cite{tagliasacchi20_interspeech}.
Configuration (B) consists of adding noise $N_{\{o,i\}}$ to the outer and in-ear microphone signals, but no interfering talkers to either sensor as in~\cite{ohlenbusch_multi-microphone_2024}.
In configuration (C), we consider training with both noise components $N_{\{o,i\}}$ and adding interferers $V_o$ only to the outer microphone signal.  
Finally, in configuration (D), we use all components during training but approximate the in-ear interferer component by an attenuated version of the outer microphone component as $V_i \approx a\cdot V_o$.

\subsection{Experimental setup}
The experiments are conducted at a sampling rate of 16\,kHz, using an STFT framework with a frame length of 32\,ms and a frame shift of 16\,ms, where a square-root Hann window is used both in analysis and synthesis. 
Training is carried out with a batch size of 8 and an example length of 3 seconds (randomly selected clip).
Mean-variance normalization based on clean speech training subset statistics is applied to each input channel independently.
For personalized systems, a single disjoint utterance $\tilde{S}$ from the same talker was randomly selected as the clean enrollment speech $\tilde{Y}_\text{enroll}=\tilde{S}_{\text{enroll}{\{o,i\}}}$ (OM or IM enrollment) for each training example.

The loss consists of the combined $L_1$ difference between clean target speech at the outer microphone $S_o$ and the estimated speech signal $\hat{S_{o}}$ in both time and STFT magnitude domains~\cite{wang_stft-domain_2023}.
Gradient clipping is used if the total gradient $L_2$-norm is greater than 10.
The ADAM optimizer~\cite{kingma_adam_2015} is used with an initial learning rate of 0.001, which is halved every five epochs until training for a total of 50 epochs, after which the system from the epoch with the best validation loss is selected.
We use the time-domain SpeakerBeam~\cite{delcroix_tdspeakerbeam_2020} architecture (4.985\,M parameters without the speaker encoder)
as a baseline system and train it with the same setups described above.

\section{Results}

For evaluation, we use the scale-invariant signal-to-distortion ratio (SI-SDR)~\cite{leRouxSDR2019}, wideband PESQ~\cite{international_telecommunications_union_itu_itu-t_2001}, and ESTOI~\cite{jensen_algorithm_2016} between the target and estimated clean speech signal at the outer microphone.
Results are averaged across the entire test set. 
For evaluation, target speech is either mixed with noise (\textbf{N}), interfering voices (\textbf{V}) or both (\textbf{N+V}).
To evaluate robustness against noisy enrollment speech in Section~\ref{sec:noisy_enrollment}, clean enrollment utterances are mixed with noise $\tilde{N}$, so $\tilde{Y}_\text{enroll}=\tilde{S}_{\text{enroll}{\{o,i\}}}+\tilde{N}_{\{o,i\}}$, and the performance of the system is evaluated only adding interfering speech (\textbf{V}) to the target speech.

\subsection{In-domain evaluation}

Table~\ref{tab:results_VV_full_SISDR_PESQ_STOI} shows the results for SE and PSE systems evaluated on the Vibravox dataset.
All systems meaningfully improve speech quality in terms of objective metrics when only noise is present. For the SE system, in the cases of interferer or interferer and noise, only slight improvements over the noisy signals are achieved, indicating the system does not have enough information to suppress interferers in these scenarios. On the other hand, enhancement using the PSE or SpeakerBeam systems yields meaningful interferer reduction. 
When comparing the outer with the in-ear microphone for enrollment, both systems are observed to benefit more from personalization with the outer microphone. 
Despite being smaller, the FT-JNF PSE systems achieve slightly higher scores than SpeakerBeam in all scenarios.

\begin{table}
 \vspace{-6pt}
\caption{Vibravox evaluation results for target speech mixed with noise (\textbf{N}), with an interferer (\textbf{V}), or with both (\textbf{N+V}). Personalization based on outer and in-ear microphone enrollment utterances is denoted by OM and IM, respectively. SB: SpeakerBeam~\cite{delcroix_tdspeakerbeam_2020}.}
\label{tab:results_VV_full_SISDR_PESQ_STOI}
\resizebox{\columnwidth}{!}{
\setlength{\tabcolsep}{0.5em}
\begin{tabular}{lc|rrr|rrr|rrr}
\toprule
& & \multicolumn{3}{c}{\textit{SI-SDR (dB)}} & \multicolumn{3}{c}{\textit{PESQ}} & \multicolumn{3}{c}{\textit{ESTOI}}\\
\textbf{System} & \textbf{Enrol.} & \textbf{N} & \textbf{V} & \textbf{N+V} & \textbf{N} & \textbf{V} & \textbf{N+V} & \textbf{N} & \textbf{V} & \textbf{N+V}\\
\midrule
 Noisy & - & 0.10 & -0.04 & -4.50 & 1.21 & 1.24 & 1.11 & 0.53 & 0.54 & 0.35 \\
\midrule
SE & - & 7.64 & 0.55 & -1.76 & 1.67 & 1.31 & 1.20 & 0.62 & 0.53 & 0.38 \\
\midrule
\multirow{2}{*}{SB \cite{delcroix_tdspeakerbeam_2020}} & OM  & 7.87 & 5.21 & 1.83 & 1.49 & 1.39 & 1.19 & 0.60 & 0.61 & 0.41 \\
 & IM & 7.64 & 4.28 & 0.82 & 1.47 & 1.35 & 1.18 & 0.59 & 0.59 & 0.39 \\
\midrule
\multirow{2}{*}{PSE} & OM  &  \textbf{8.72} & \textbf{5.78} & \textbf{2.65} & \textbf{1.74} & \textbf{1.56}& \textbf{1.31} & \textbf{0.65} & \textbf{0.64} & \textbf{0.47} \\

 & IM & 8.26 & 4.81 & 1.77 & 1.72 & 1.52 & 1.29 & 0.64 & 0.63 & 0.45 \\
\bottomrule
\end{tabular}
}
\end{table}

Table~\ref{tab:vv_2ch_results} shows the results for noise reduction by AS-SE and PAS-SE systems trained with different configurations of noise and interferer in-ear components evaluated on the Vibravox dataset.
When no in-ear noise or interferer components are used (A), performance is poor, but including in-ear noise in training (B, C, D) yields large improvements, highlighting the importance of modeling the noise leakage to the in-ear microphone during training. 
As expected for noise reduction, different configurations of interferer in-ear components do not lead to substantial performance differences between configurations (B), (C), and (D).
Similarly, added personalization (PAS-SE) does not yield improvements. 
\begin{table}
 \vspace{-6pt}
\caption{Vibravox evaluation results of FT-JNF-based AS-SE and PAS-SE systems optimized with different training configurations. Only noise suppression is evaluated, since the Vibravox dataset does not contain in-ear microphone signals of isolated interfering voices.
}
\label{tab:vv_2ch_results}
\resizebox{\columnwidth}{!}{
\setlength{\tabcolsep}{0.5em}
\begin{tabular}{lcl|cccc|rrr}
\toprule
 &  &  & \multicolumn{4}{c|}{Configuration}  & \textit{SI-SDR (dB)}     &\textit{PESQ}     & \textit{ESTOI} \\ 
\textbf{System} & \textbf{Enrol.} & \textbf{Train.} & $N_o$ & $V_o$&$N_i$ &$ V_i$ & \textbf{N} & \textbf{N} & \textbf{N} \\
\midrule
 Noisy & - & & \multicolumn{4}{c|}{N/A} & 0.10 & 1.21 & 0.53 \\
\midrule
\multirow{5}{*}{AS-SE} & - & A \cite{tagliasacchi20_interspeech} &\cmark&\cmark&\xmark&\xmark& -0.92 & 1.25 & 0.50 \\
\cmidrule{2-10}
  & - & B \cite{ohlenbusch_multi-microphone_2024} &\cmark&\xmark&\cmark&\xmark&             \textbf{11.23} & \textbf{2.19} & \textbf{0.75} \\
\cmidrule{2-10}
  & - & C & \cmark & \cmark & \cmark & \xmark & 
10.42 & 2.06 & 0.73 \\
\cmidrule{2-10}
  & - & D     & \cmark & \cmark &\cmark & $a\cdot V_o$ &       9.98  & 2.02 & 0.72 \\
\midrule
\multirow{5}{*}{PAS-SE} & OM &\multirow{2}{*}{C} & \multirow{2}{*}{\cmark}&\multirow{2}{*}{\cmark}&\multirow{2}{*}{\cmark}&\multirow{2}{*}{\xmark} &             10.68 & 2.09 & 0.73 \\
 & IM & & &&& &             10.88 & 2.14 & 0.74 \\
\cmidrule{2-10}
 &                         OM &\multirow{2}{*}{D} & \multirow{2}{*}{\cmark}&\multirow{2}{*}{\cmark}&\multirow{2}{*}{\cmark}&\multirow{2}{*}{$a\cdot V_o$} &       10.29 & 2.07 & 0.72 \\
 &                         IM & & &&& &       10.29 & 2.06 & 0.72 \\ 
\bottomrule
\end{tabular}
}
\end{table}

\subsection{Cross-dataset generalization}
\label{sec:cross_corpus_eval}
\begin{table*}
 \vspace{-6pt}
\caption{Oldenburg evaluation results for target speech mixed with noise (\textbf{N}), with an interferer (\textbf{V}), and with both interferer and noise (\textbf{N+V}). 
All models trained on Vibravox data, except those denoted by OL where the system is trained on the in-domain Oldenburg training dataset.
}
\label{tab:results_OL_full}
\centering
\resizebox{\textwidth}{!}{
\renewcommand{\arraystretch}{.75}
\begin{tabular}{lcl|cccc|rrr|rrr|rrr}
\toprule
& & & \multicolumn{4}{c}{Configuration} &\multicolumn{3}{c}{\textit{SI-SDR (dB)}} & \multicolumn{3}{c}{\textit{PESQ}} & \multicolumn{3}{c}{\textit{ESTOI}}\\
\textbf{System} & \textbf{Enrol.} & \textbf{Train.} & $N_o$ & $V_o$&$N_i$ &$ V_i$& \textbf{N} & \textbf{V} & \textbf{N+V} & \textbf{N} & \textbf{V} & \textbf{N+V} & \textbf{N} & \textbf{V} & \textbf{N+V}\\
\midrule
Noisy & &-& \multicolumn{4}{c|}{\multirow{8}{*}{N/A}} & 0.13 & 0.04 & -4.52 & 1.28 & 1.51 & 1.18 & 0.40 & 0.55 & 0.30\\ 
\cmidrule{1-3} \cmidrule{8-16}
SE &-& - &&& & & 8.04 & 2.23 & -0.06 & 1.60 & 1.58 & 1.31 & 0.46 & 0.57 & 0.34 \\
\cmidrule{1-3} \cmidrule{8-16}
\multirow{2}{*}{SpeakerBeam \cite{delcroix_tdspeakerbeam_2020}} & OM& - &&&&& -1.42 & -7.02 & -8.63 & 1.19 & 1.24 & 1.13 & 0.28 & 0.39 & 0.19\\
                     & IM& - &&&&& -1.64 & -5.21 & -7.44 & 1.16 & 1.20 & 1.11 & 0.28 & 0.38 & 0.19\\
\cmidrule{1-3} \cmidrule{8-16}
\multirow{2}{*}{PSE}    & OM& - &&&&& 8.29 & 4.67 & 2.01 & 1.63 & 1.70 & 1.36 & 0.47 & 0.58 & 0.36\\
             & IM& - &&&&& 8.43 & 4.45 & 2.00 & 1.61 & 1.67 & 1.35 & 0.47 & 0.58 & 0.37 \\
\midrule
\multirow{6}{*}{AS-SE} & - & A \cite{tagliasacchi20_interspeech} &\cmark&\cmark&\xmark&\xmark& 3.35 & 3.52 & -1.72 & 1.51 & 1.78 & 1.33 & 0.49 & 0.63 & 0.41\\
\cmidrule{2-16}
      & - &B \cite{ohlenbusch_multi-microphone_2024} &\cmark&\xmark&\cmark&\xmark& 9.85 & 2.62 & 2.60 & 1.87 & 1.67 & 1.50 & 0.56 & 0.60 & 0.46\\
\cmidrule{2-16}
     & - & C  & \cmark & \cmark & \cmark & \xmark & 10.09 & 5.15 & 3.59 & 1.88 & 1.90 & 1.57 & 0.56 & 0.64 & 0.48\\
\cmidrule{2-16}
      & - & D & \cmark & \cmark & \cmark & $a\cdot V_o$ & 9.63 & 7.20 & 4.97 & 1.86 & 1.93 & 1.60 & 0.55 & 0.64 & 0.47\\
\midrule 
\multirow{5}{*}{PAS-SE}       & OM&\multirow{2}{*}{C} & \multirow{2}{*}{\cmark}&\multirow{2}{*}{\cmark}&\multirow{2}{*}{\cmark}&\multirow{2}{*}{\xmark}   & 10.57 & 6.38 & 4.87 & 1.92 & 1.95 & 1.62 & 0.56 & 0.65 & 0.48\\
     & IM& &&&&   & \textbf{10.70} & 7.40 & 5.35 & 1.92 & 2.01 & 1.64 & \textbf{0.57} & \textbf{0.66} & \textbf{0.49}\\
\cmidrule{2-16}
     & OM& \multirow{2}{*}{D} &\multirow{2}{*}{\cmark}&\multirow{2}{*}{\cmark}&\multirow{2}{*}{\cmark}&\multirow{2}{*}{$a\cdot V_o$}& 9.24 & 7.17 & 5.31 & 1.85 & 1.89 & 1.61 & 0.54 & 0.64 & 0.48\\
     & IM& &&&&& 10.30 & \textbf{8.34} & \textbf{5.85} & 1.88 & 1.98 & 1.62 & 0.55 & 0.65 & 0.48\\
\hline
\midrule
AS-SE    & - & OL &\multicolumn{4}{c|}{N/A}& 7.47 & 7.27 & 4.57 & 1.88 & 2.08 & 1.72 & 0.51 & 0.63 & 0.46\\
\midrule 
\multirow{2}{*}{PAS-SE}     & OM& \multirow{2}{*}{OL}&\multicolumn{4}{c|}{\multirow{2}{*}{N/A}}& 7.45 & 7.49 & 4.64 & 1.97 & 2.19 & 1.82 & 0.53 & 0.64 & \textbf{0.49}\\
     & IM& &&&&& 7.63 & 7.57 & 4.85 & \textbf{2.00} & \textbf{2.23} & \textbf{1.84} & 0.53 & 0.64 & \textbf{0.49}\\
 \bottomrule
\end{tabular}
}
\end{table*}
Table~\ref{tab:results_OL_full} shows the cross-dataset generalization (out-of-domain) results of various
Vibravox-trained systems evaluated on the Oldenburg dataset. Again, the non-personalized SE system is able to reduce noise, but achieves little improvement in terms of reducing interfering talkers.
The PSE baseline SpeakerBeam trained on Vibravox fails to enhance speech on the Oldenburg dataset, exhibiting SI-SDR scores below that of noisy signals, while FT-JNF-based PSE trained in the same way achieves similar scores on both Oldenburg and Vibravox test sets. This may be due to the fact that SpeakerBeam operates in the time domain, employing learnable filterbanks, which may be more prone to dataset-specific biases, while FT-JNF uses magnitude STFT features.

For the AS-SE system trained considering noise and interferer only at the outer microphone (A), performance is unsurprisingly poor, due to the fact that in-ear noise or interferers were not present during training. 
In configuration (B), where the system is only trained with environmental noise and without interferers, performance for noise reduction improves, whereas interferer reduction does not. When an interfering talker is added only at the outer microphone along with noise at both sensors (C), both noise and interferer reduction performance improves, indicating that interferer modeling at the in-ear microphone may not be strictly necessary to enable interferer rejection.
However, the best performance is achieved when in-ear interferers are approximated and added to recorded in-ear noise (D). 

In the PAS-SE systems, we observe systematic performance improvements compared to their AS-SE counterparts, especially on but not limited to interferers. 
For a PAS-SE system trained with configuration (C), interferer suppression improves by conditioning with either outer or in-ear microphone enrollments, indicating the conditioning may compensate for the absence of interfering voice leakage to the in-ear microphone during training. 
For configuration (D), further improvements are observed compared to (C) in SI-SDR for interferers and noise and interferer, while other metrics remain similar.
Interestingly, for systems both trained and evaluated on the Oldenburg data (OL), personalization yields smaller improvements, likely due to only 12 speakers being available in the training subset. 

These performance trends indicate that the AS-SE systems require the use of some level of interferer signals during training (C,~D) to generalize to an unseen auxiliary-sensor array. Cross-domain performance improves more when combined with enrollment-based personalization (PAS-SE), nearing or even outperforming the corresponding matched system trained on in-domain data (OL).
\subsection{Personalization with noisy enrollment signals}
\label{sec:noisy_enrollment}
\begin{figure}
    \centering
    \includegraphics
    [width=0.9\columnwidth]
{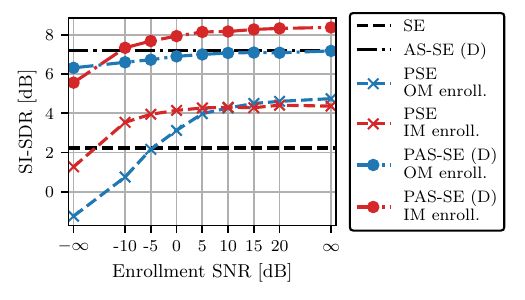}
    \vspace{-0.5cm}
    \caption{
    Cross-dataset interferer reduction performance (\textbf{V}) achieved by SE, PSE, AS-SE, and PAS-SE systems at different enrollment utterance SNRs ($-\infty$: only noise, no speech, $\infty$: clean speech).
    }
    \label{fig:noisy_enrollment_plot}
\end{figure}

The performance of the proposed systems personalized with either in-ear or outer microphone enrollment utterances was evaluated with noisy enrollment signals~(Fig.~\ref{fig:noisy_enrollment_plot}). Similar to Table~\ref{tab:results_OL_full}, the systems were trained on Vibravox and evaluated on Oldenburg datasets (out-of-domain). For PAS-SE and AS-SE, we use the model trained with configuration (D), which yielded the overall best results.

For PSE systems, conditioning with an in-ear microphone is more robust to noise in the enrollment utterance, likely due to the in-ear microphone being acoustically shielded from noise. Using the in-ear microphone yields improvements over the non-personalized single-channel system even at -10\,dB enrollment SNR, whereas using the outer microphone for conditioning provides no benefits below 0\,dB.
Similarly, the PAS-SE systems conditioned with the in-ear enrollment speech achieve better performance than the AS-SE systems for enrollment SNRs higher than -10\,dB, while PAS-SE systems conditioned with an outer microphone do not benefit from personalization at all. While not considered here, training with enrollment augmentation as in~\cite{li_effectiveness_2024} may provide further benefits. 

\vspace{-0.3cm}


\section{Conclusion}
In this paper, we have systematically evaluated PSE and AS-SE in terms of denoising, interferer suppression, and cross-dataset generalization. Experimental results demonstrate that the proposed training-time augmentations can yield substantial cross-dataset generalization benefits for AS-SE systems, nearing the performance of systems trained fully in-domain. Moreover, the proposed PAS-SE systems with enrollment-based personalization generalize across datasets, outperforming dataset-specific baselines, and retain their benefits with in-ear enrollments even when the enrollment contains noise.
We hope that the proposed methodology can facilitate future research in the direction of fully device-agnostic AS-SE systems.
\vfill
\pagebreak

\footnotesize
\bibliographystyle{IEEEbib}
\bibliography{references}

\begin{thebibliography}{10}

\bibitem{braun_towards_2021}
Sebastian Braun, Hannes Gamper, Chandan~K.A. Reddy, and Ivan Tashev,
\newblock ``Towards efficient models for real-time deep noise suppression,''
\newblock in {\em Proc. International Conference on Acoustics, Speech and Signal Processing (ICASSP)}, 2021, pp. 656--660.

\bibitem{rong_gtcrn_2024}
Xiaobin Rong, Tianchi Sun, Xu~Zhang, Yuxiang Hu, Changbao Zhu, and Jing Lu,
\newblock ``{GTCRN}: A speech enhancement model requiring ultralow computational resources,''
\newblock in {\em Proc. International Conference on Acoustics, Speech and Signal Processing (ICASSP)}, 2024, pp. 971--975.

\bibitem{fedorov2020tinylstms}
Igor Fedorov, Marko Stamenovic, Carl Jensen, Li-Chia Yang, Ari Mandell, Yiming Gan, Matthew Mattina, and Paul~N Whatmough,
\newblock ``{TinyLSTMs}: Efficient neural speech enhancement for hearing aids,''
\newblock in {\em Proc. Interspeech}, 2020, pp. 4054--4058.

\bibitem{nathoo2024two}
Rayan~Daod Nathoo, Mikolaj Kegler, and Marko Stamenovic,
\newblock ``Two-step knowledge distillation for tiny speech enhancement,''
\newblock in {\em Proc. International Conference on Acoustics, Speech and Signal Processing (ICASSP)}, 2024, pp. 10141--10145.

\bibitem{ChangListening2016}
Cheng-Yuan Chang, Antonius Siswanto, Chung-Ying Ho, Ting-Kuo Yeh, Yi-Rou Chen, and Sen~M. Kuo,
\newblock ``Listening in a noisy environment: Integration of active noise control in audio products,''
\newblock {\em IEEE Consumer Electronics Magazine}, vol. 5, no. 4, pp. 34--43, 2016.

\bibitem{bouserhal_-ear_2019}
Rachel~E. Bouserhal, Antoine Bernier, and J{\'e}r{\'e}mie Voix,
\newblock ``{An In-Ear Speech Database in Varying Conditions of the Audio-Phonation Loop},''
\newblock {\em J. Acoust. Soc. Am.}, vol. 145, no. 2, pp. 1069--1077, 2019.

\bibitem{ohlenbusch_modeling_2023}
Mattes Ohlenbusch, Christian Rollwage, and Simon Doclo,
\newblock ``Modeling of speech-dependent own voice transfer characteristics for hearables with an in-ear microphone,''
\newblock {\em Acta Acustica}, vol. 8, 2024.

\bibitem{gaardbaek_origin_2025}
Bjarke Gårdbæk and Preben Kidmose,
\newblock ``On the origin of cardiovascular sounds recorded from the ear,''
\newblock {\em IEEE Trans. Biomedical Engineering}, vol. 72, no. 1, pp. 210--216, 2025.

\bibitem{tagliasacchi20_interspeech}
Marco Tagliasacchi, Yunpeng Li, Karolis Misiunas, and Dominik Roblek,
\newblock ``{SEANet}: A multi-modal speech enhancement network,''
\newblock in {\em Proc. Interspeech}, Oct. 2020, pp. 1126--1130.

\bibitem{yu_time-domain_2020}
Cheng Yu, Kuo-Hsuan Hung, Syu-Siang Wang, Yu~Tsao, and Jeih-Weih Hung,
\newblock ``{Time-Domain Multi-Modal Bone/Air Conducted Speech Enhancement},''
\newblock {\em IEEE Signal Processing Letters}, vol. 27, pp. 1035--1039, 2020.

\bibitem{wang_fusing_2022}
Heming Wang, Xueliang Zhang, and DeLiang Wang,
\newblock ``{Fusing Bone-Conduction and Air-Conduction Sensors for Complex-Domain Speech Enhancement},''
\newblock {\em IEEE/ACM Trans. on Audio, Speech, and Language Processing}, vol. 30, pp. 3134--3143, 2022.

\bibitem{wang_multi-modal_2022}
Mou Wang, Junqi Chen, Xiaolei Zhang, Zhiyong Huang, and Susanto Rahardja,
\newblock ``{Multi-Modal Speech Enhancement with Bone-Conducted Speech in Time Domain},''
\newblock {\em Applied Acoustics}, vol. 200, no. 109058, 2022.

\bibitem{edraki_speaker_2024}
Amin Edraki, Wai-Yip Chan, Jesper Jensen, and Daniel Fogerty,
\newblock ``Speaker {{Adaptation For Enhancement Of Bone-Conducted Speech}},''
\newblock in {\em Proc. {{International Conference}} on {{Acoustics}}, {{Speech}} and {{Signal Processing}} ({{ICASSP}})}, 2024, pp. 10456--10460.

\bibitem{ohlenbusch_multi-microphone_2024}
Mattes Ohlenbusch, Christian Rollwage, and Simon Doclo,
\newblock ``{Mul\-ti\--Mi\-cro\-phone} {Noise} {Data} {Augmentation} for {{DNN-based}} {Own Voice Reconstruction} for {Hearables} in {Noisy} {Environments},''
\newblock in {\em Proc. {{International Conference}} on {{Acoustics}}, {{Speech}} and {{Signal Processing}} ({{ICASSP}})}, 2024, pp. 416--420.

\bibitem{delcroix_tdspeakerbeam_2020}
Marc Delcroix, Tsubasa Ochiai, Katerina Zmolikova, Keisuke Kinoshita, Naohiro Tawara, Tomohiro Nakatani, and Shoko Araki,
\newblock ``Improving speaker discrimination of target speech extraction with time-domain speakerbeam,''
\newblock in {\em Proc. International Conference on Acoustics, Speech and Signal Processing (ICASSP)}, 2020, pp. 691--695.

\bibitem{Xu_Spex_2020}
Chenglin Xu, Wei Rao, Eng~Siong Chng, and Haizhou Li,
\newblock ``{SpEx}: Multi-scale time domain speaker extraction network,''
\newblock {\em IEEE/ACM Trans. on Audio, Speech, and Language Processing}, vol. 28, pp. 1370--1384, 2020.

\bibitem{eskimez_personalized_2022}
Sefik~Emre Eskimez, Takuya Yoshioka, Huaming Wang, Xiaofei Wang, Zhuo Chen, and Xuedong Huang,
\newblock ``Personalized speech enhancement: new models and comprehensive evaluation,''
\newblock in {\em Proc. International Conference on Acoustics, Speech and Signal Processing (ICASSP)}, 2022, pp. 356--360.

\bibitem{ZmolikovaOverview2023}
Katerina Zmolikova, Marc Delcroix, Tsubasa Ochiai, Keisuke Kinoshita, Jan Černocký, and Dong Yu,
\newblock ``Neural target speech extraction: An overview,''
\newblock {\em IEEE Signal Processing Magazine}, vol. 40, no. 3, pp. 8--29, 2023.

\bibitem{sinha2024variants}
Ragini Sinha, Christian Rollwage, and Simon Doclo,
\newblock ``Variants of {LSTM} cells for single-channel speaker-conditioned target speaker extraction,''
\newblock {\em EURASIP Journal on Audio, Speech, and Music Processing}, vol. 2024, no. 1, pp. 63, 2024.

\bibitem{tesch_insights_2023}
Kristina Tesch and Timo Gerkmann,
\newblock ``Insights {Into} {Deep} {Non}-{Linear} {Filters} for {Improved} {Multi}-{Channel} {Speech} {Enhancement},''
\newblock {\em IEEE/ACM Trans. on Audio, Speech, and Language Processing}, vol. 31, pp. 563--575, 2023.

\bibitem{ohlenbusch_lowcomplexity_2025}
Mattes Ohlenbusch, Christian Rollwage, and Simon Doclo,
\newblock ``Low-complexity own voice reconstruction for hearables with an in-ear microphone,''
\newblock in {\em Proc. International Conference on Acoustics, Speech and Signal Processing (ICASSP)}, 2025.

\bibitem{pariente20_interspeech}
Manuel Pariente, Samuele Cornell, Joris Cosentino, Sunit Sivasankaran, Efthymios Tzinis, Jens Heitkaemper, Michel Olvera, Fabian-Robert Stöter, Mathieu Hu, Juan~M. Martín-Doñas, David Ditter, Ariel Frank, Antoine Deleforge, and Emmanuel Vincent,
\newblock ``{Asteroid}: The {PyTorch}-based audio source separation toolkit for researchers,''
\newblock in {\em Proc. Interspeech}, 2020, pp. 2637--2641.

\bibitem{jhauret-et-al-2024-vibravox}
Julien Hauret, Malo Olivier, Thomas Joubaud, Christophe Langrenne, Sarah Poir{\'e}e, Véronique Zimpfer, and {\'E}ric Bavu,
\newblock ``{Vibravox: A Dataset of French Speech Captured with Body-conduction Audio Sensors},''
\newblock {\em Speech Communication}, vol. 172, pp. 103238, 2025.

\bibitem{denk_one-size-fits-all_2019}
Florian Denk, Miriam Lettau, Henning Schepker, Simon Doclo, Reinhild Roden, Matthias Blau, Jörg-Hendrik Bach, Jan Wellmann, and Birger Kollmeier,
\newblock ``A one-size-fits-all earpiece with multiple microphones and drivers for hearing device research,''
\newblock in {\em Proc. {AES} {International} {Conference} on {Headphone} {Technology}}, 2019.

\bibitem{dubey_icassp_2024}
Harishchandra Dubey, Ashkan Aazami, Vishak Gopal, Babak Naderi, Sebastian Braun, Ross Cutler, Alex Ju, Mehdi Zohourian, Min Tang, Mehrsa Golestaneh, and Robert Aichner,
\newblock ``{{ICASSP}} 2023 {{Deep Noise Suppression Challenge}},''
\newblock {\em IEEE Open Journal of Signal Processing}, vol. 5, pp. 725--737, 2024.

\bibitem{wang_stft-domain_2023}
Zhong-Qiu Wang, Gordon Wichern, Shinji Watanabe, and Jonathan Le~Roux,
\newblock ``{{STFT-domain}} neural speech enhancement with very low algorithmic latency,''
\newblock {\em IEEE/ACM Trans. on Audio, Speech, and Language Processing}, vol. 31, pp. 397--410, 2023.

\bibitem{kingma_adam_2015}
Diederik~P. Kingma and Jimmy Ba,
\newblock ``Adam: {{A}} method for stochastic optimization,''
\newblock in {\em Proc. {{International Conference on Learning Representations}}}, 2015.

\bibitem{leRouxSDR2019}
Jonathan {Le Roux}, Scott Wisdom, Hakan Erdogan, and John~R. Hershey,
\newblock ``{SDR} – half-baked or well done?,''
\newblock in {\em Proc. International Conference on Acoustics, Speech and Signal Processing (ICASSP)}, 2019, pp. 626--630.

\bibitem{international_telecommunications_union_itu_itu-t_2001}
{International Telecommunications Union (ITU)},
\newblock ``{ITU}-{T} {P}.862, {Perceptual} evaluation of speech quality ({PESQ}): {An} objective method for end-to-end speech quality assessment of narrow-band telephone networks and speech codecs,''
\newblock {\em International Telecommunications Union}, 2001.

\bibitem{jensen_algorithm_2016}
Jesper Jensen and Cees~H. Taal,
\newblock ``An {{Algorithm}} for {{Predicting}} the {{Intelligibility}} of {{Speech Masked}} by {{Modulated Noise Maskers}},''
\newblock {\em IEEE/ACM Trans. on Audio, Speech, and Language Processing}, vol. 24, no. 11, pp. 2009--2022, 2016.

\bibitem{li_effectiveness_2024}
Junjie Li, Ke~Zhang, Shuai Wang, Haizhou Li, Man-Wai Mak, and Kong~Aik Lee,
\newblock ``On the effectiveness of enrollment speech augmentation for target speaker extraction,''
\newblock in {\em IEEE Spoken Language Technology Workshop (SLT)}, 2024, pp. 325--332.

\end{thebibliography}

\end{document}